\documentclass[aps,pra,amsmath,amssymb,amsfonts,showpacs,superscriptaddress,floatfix,twocolumn]{revtex4} 
\usepackage{amsmath}
\usepackage{graphicx}
\usepackage{amsthm}
\usepackage{color}

\newcommand{\be}{\begin{equation}}
\newcommand{\ee}{\end{equation}}

\begin{document}
\title{Towards Practical Quantum Variational Algorithms}

\author{Dave Wecker}
\affiliation{Quantum Architectures and Computation Group, Microsoft Research, Redmond, WA 98052, USA}

\author{Matthew~B.~Hastings}
\affiliation{Station Q, Microsoft Research, Santa Barbara, CA 93106-6105, USA}
\affiliation{Quantum Architectures and Computation Group, Microsoft Research, Redmond, WA 98052, USA}

\author{Matthias Troyer}
\affiliation{Theoretische Physik, ETH Zurich, 8093 Zurich, Switzerland}
\begin{abstract}
The preparation of quantum states using short quantum circuits is one of the most promising near-term applications of small quantum computers, especially if the circuit is short enough and the fidelity of gates high enough that it can be executed without quantum error correction. Such quantum state preparation can be used in variational approaches, optimizing parameters in the circuit to minimize the energy of the constructed quantum state for a given problem Hamiltonian. For this purpose we propose a simple-to-implement class of quantum states motivated by adiabatic state preparation. We test its accuracy and determine the required circuit depth for a Hubbard model on ladders with up to 12 sites (24 spin-orbitals), and for small molecules. We find that this ansatz converges faster than previously proposed schemes based on unitary coupled clusters. While the required number of measurements is astronomically large for quantum chemistry applications to molecules, applying the variational approach to the Hubbard model (and related models) is found to be far less demanding and potentially practical on small quantum computers.
We also discuss another application of quantum state preparation using short quantum circuits, to prepare trial ground states of models faster than using adiabatic state preparation. 
\end{abstract}
\maketitle

\section{Introduction}

Variational classes of states
such as matrix product states (MPS)  \cite{mps}, multiscale entanglement renormalization (MERA)  \cite{mera}, and projected entangled pair states (PEPS)  \cite{peps} play a key role in studying many-body quantum systems. An ideal class of
states should be large enough to approximate the ground state and it must be possible to evaluate
the energy and observables. Unfortunately for many potential classes of states, including PEPS, evaluation may be difficult on a classical computer, and for other classes it can be very computationally expense. However, on a {\em quantum computer}, large classes of PEPS states can be prepared efficiently  \cite{pepsprep1,pepsprep2}, (although likely not all PEPS can be prepared this way  \cite{schuch}). Once the state is prepared, the energy and observables
can be measured by sampling using short quantum circuits. Similar quantum circuits exist for MPS and MERA states  \cite{univary}. 
Recently, Ref.~\onlinecite{aag} proposed variational methods for studying quantum chemistry problems on a quantum computer, and
demonstrated one method on a model with a four-dimensional Hilbert space. 

Preparation of variational states is an attractive application of small quantum computers, since many classically intractable states can be prepared with quite short quantum circuits and thus do not pose stringent requirements on coherence times and gate fidelities. While variation over all possible states that can be produced by quantum circuits of a given maximum depth (as done in the demonstration experiments of Ref.~\onlinecite{aag}) might sound ideal, this approach does not scale efficiently. Ref.~\onlinecite{aag} also proposed using the unitary coupled cluster method (UCC)  \cite{ucc}, where variational
states are prepared using unitary evolution under a sum of fermionic terms including  $c^\dagger_p c_q+h.c.$, $c^\dagger_p c^\dagger_q c_r c_s+h.c.$, and higher order terms, typically
restricted to the case that creation operators are on unoccupied orbitals in the Hartree-Fock state and annihilation operators are on occupied (or vice-versa as required by hermiticity).
That approach involves many parameters: even if truncated
at the level of four fermion operators, the number of variational parameters scales as the number of occupied orbitals squared times the number of unoccupied orbitals squared, which at
constant filling fraction
is the fourth power of the number of orbitals.

In this paper, we present an analysis of a different class of variational states, that we term ``Hamiltonian variational", using a modest gate depth and a very modest number of variational parameters compared to the system size. We present a detailed numerical analysis of this variational technique applied to a Hubbard model, ranging up to systems of $12$ sites (equivalently, $24$ spin-orbitals) at half-filling, to quantify its accuracy. The difficulty in general increases with system size, but also fluctuates in complicated ways, depending upon the spectrum of the non-interacting model. For certain cases, including our largest size, the non-interacting model is highly
degenerate, leading to strong interaction effects and a very poor overlap of an initial Slater determinant state with the true ground state. We find that even then the variational approach works well. Finally, we also analyze the effects of sampling error in measuring energy and give concrete estimates for time scales to implement this procedure in practice.

The Hamiltonian variational method builds the variational state using rotations by terms in the Hamiltonian.  This is then very well suited for models such as the Hubbard model where
there are few interactions terms, all of which take a simple form; in contrast, for UCC applied to the Hubbard model, the large number of possible terms 
$c^\dagger_p c^\dagger_q c_r c_s+h.c.$ will lead to a much larger circuit depth and further each term will take a more complicated form than the simple interaction in the Hubbard model.  So, for comparison purposes, we instead apply both methods to several quantum chemistry systems also where the number of terms in the Hamiltonian is comparable to the number of possible four fermion terms.  In this case, we consider additionally several variants of UCC that we term Rxx (different variants include different terms) described below and we show that it is possible to improve the UCC circuit depth by taking a very limited Trotter number.
While UCC shows a high accuracy on small molecules, we find that this requires more evaluations.  Further, on larger molecules with stronger interaction, we find that
the accuracy of UCC and Rxx is worse than that that obtained by the Hamiltonian variational method.


The class of variational states that we consider is inspired by both adiabatic state preparation and the quantum optimization algorithm of Refs.~\onlinecite{farhi1,farhi2}.
 Consider a family of Hamiltonians $H=\sum_a J_a h_a$ with the $J_a$ being scalars and the $h_a$ being operators. 
Let $J_a^0$ and $J_a^1$ be two choices of these $J_a$, with corresponding Hamiltonians $H_0,H_1$.
Suppose it is easy to prepare a ground state $\Psi_I$ of $H_0$ (for example, in our study of the Hubbard model, $H_0$ is non-interacting). Then, assuming that no gap closes, it is possible to adiabatically evolve from $\Psi_I$ to the ground state of $H_1$. If we break the annealing into short time steps ${\rm d}t$ and evolve for a total time $T$, this annealing is a sequence of $(T/{\rm d}t)$ different unitary rotations by Hamiltonians interpolating between $H_0$ and $H_1$.
This could be implemented on a quantum computer using a Trotter-Suzuki method which further decomposes this sequence into a sequence of unitary rotations by individual terms in the Hamiltonian.

The idea of the variational method that we consider is still restricted to a sequence of unitary rotations by terms in the Hamiltonian, but considers arbitrary angles for the rotations in the
sequence, rather than choosing them from a Trotterization of an annealing process. By choosing these angles arbitrarily, this allows us to take a much shorter sequence.
Thus, we consider a trial state, which we call the ``Hamiltonian variational'' state
\be
\label{VarEq}
\Psi_T=\exp(i \theta_n h_{a_n}) \ldots \exp(i \theta_2 h_{a_2}) \exp(i \theta_1 h_{a_1})
\Psi_I,
\ee
where larger $n$ increases the accuracy. The angles $\theta_k$ are variational parameters.
The optimziation algorithm of Refs.~\onlinecite{farhi1,farhi2} considered only rotation by two different types of operators, while we consider rotation by a larger number of terms.

If all the terms in the Hamiltonian have some symmetry (such as spin-rotation symmetry), then $\Psi_I$ and $\Psi_T$ transform in the same way under this symmetry.
This allows us to find ground states with different quantum numbers by picking initial states $\Psi_I$ with the desired quantum number.
In cases of symmetries such as translation invariance, it is helpful to try to construct the sequence of terms $a_k$ to approximately preserve this symmetry (see a more detailed discussion below).

To be useful for variational state preparation, it must be possible to optimize over the given parameters, without getting stuck in false minima. In Sec. \ref{sec:exact} we discuss techniques which succeed in finding good optima for the Hubbard model using access to numerically exact values of the energy obtained from a classical simulation. Such a classical approach may be useful on a quantum computer as it finds short circuits that prepare specific highly entangled states, albeit limited to system sizes where classical simulation is possible. 

To go beyond the limits of classical simulation on a quantum computer one must measure the energy using a quantum circuit. One way is to write the Hamiltonian as a sum of sets of terms, so that all terms in a given set commute with each other and then measure each set in one run, doing many runs for each set of terms, with the error in the energy going as the inverse square-root of the number of samples. The required number of samples to achieve high accuracy may be large and limiting the number of samples significantly impacts the ability to find the optimum as discussed later in Sec. \ref{sec:inexact}.

Another way to measure energy is to to use phase estimation to compute the energies, rather than sampling. In this case, the error in energy goes inversely with the phase estimation time. This procedure still is probabilistic, returning on each run an energy chosen randomly from a distribution. Thus many runs
are still required to estimate the average energy. For many distributions, this procedure provides much more acccurate access to the energy for a given run time than
sampling would, at the cost of requiring longer coherence time. Additionally, it allows one to optimize on the probability of finding the ground state, rather
than on the energy, which may improve the optimization. 

 One may wonder why this would be useful, since if phase estimation finds an energy equal to the ground state, one knows that one has successfully prepared the ground state and can now make a measurement, without any need to improve the variational state. However, once the optimum variational parameters are found, these parameters are classical information that can be used to quickly re-create the state. This is useful if one wishes to make many measurements on the state, if the measurements destroy the state. Also, techniques in Ref.~\onlinecite{hubbard} show how to quadratically speed up the sampling of properties of the state, assuming access by measuring a projector onto the state. However, if we have access to a projector onto $\Psi_I$, then conjugating this projector by the unitaries in Eq.~(\ref{VarEq}) gives us a short-depth circuit which projects
$\Psi_T$.
Finally, in many applications such as chemical reactions where one studies the same Hamiltonian along a path of parameters, it may be possible to use the
variational solution for a given Hamiltonian to find a solution of a nearby Hamiltonian rapidly.

\section{The Hubbard model}

\subsection{The Hubbard Hamiltonian}

To study performance scaling across a range of sizes, we consider a sequence of Hubbard models on $N$-sites, with the sites arranged in a two-leg ladder (an $N/2$-by-$2$ square lattice) for $N=4,6,8,10,12$. The Hamiltonian of the Hubbard model is
\be
H=-t \sum_{\langle i,j \rangle} \sum_\sigma t_{ij} c^\dagger_{i,\sigma} c_{j,\sigma} + U \sum_i c^\dagger_{i,\uparrow} c_{i,\uparrow} c^\dagger_{i,\downarrow} c_{i,\downarrow},
\ee
with $t=1, U=2$. Here $c^\dagger_{i,\sigma}$ and $c{i,\sigma}$ create and annihilate an electron at site $i$ with spin $\sigma$ respectively. We use periodic boundary conditions along the long ``horizontal" direction of the lattice which is $N/2$ sites long and open boundary conditions in the short ``vertical'' direction so that there are a total of $N$ bonds in the horizontal direction and $N/2$ in the vertical direction, all with the same strength. We write this as $H=h_h+h_v+h_U$, where $h_h$ is the sum of hopping terms in the horizontal direction, $h_v$ is the sum of hopping terms in the vertical direction, and $h_U$ is the repulsion term.

\subsection{Ground state degeneracies and initial states}

We studied the system at half-filling with a $N$ electrons on $N$ sites. For an equal number of $N/2$ up and down electrons the single particle spectrum for $N=4,6,10$ has $N/2-1$ states below the Fermi energy and is doubly degenerate at the Fermi energy, giving a $2^2=4$-fold degenerate many-body ground state for the non-interacting ($U=0$) Hamiltonian.. For $N=8$ the $U=0$ Hamiltonian has a unique ground state, with $N/2$ states below the Fermi energy and an excitation gap. For $N=12$, there are $N/2-2$ single particle states below the Fermi energy, and four states at the Fermi energy, so that for $N/2$ up electrons and $N/2$ down electrons, the $U=0$ ground state is ${{4}\choose{2}}^2=36$-fold degenerate. These different degeneracies of the non-interacting problem impact the difficulty of solving it. They also reduce the overlap of a Slater determinant with the true ground state.
The degeneracies at $N=4$ and $12$ are due to the special choice of the vertical coupling; for generic vertical coupling, degeneracies are only seen for $N=4n+2=6,10,14,\ldots$, where $n$ is a positive integer.

The spin of the ground state at $U=2$ also depends upon $N$. It is a singlet for $N=4n=4,8,12,\ldots$ and is a triplet for $N=4n+2=6,10,\ldots$.
All the results that we report below are for $\Psi_I$ in the correct spin sector. In the case of $N=6,10$ we did this by choosing $N/2+1$ up electrons and
$N/2-1$ down electrons. In that sector, the free fermion ground state becomes non-degenerate.
Otherwise, we chose $N/2$ up electrons and $N/2$ down electrons.
To choose a unique ground state for $N=4,12$ we prepared the ground state of the Hamiltonian $th_h+(1-\epsilon)th_v$ for small $\epsilon>0$; this state is independent of $\epsilon$ for $\epsilon$ small and this simply picks out one of the ground states of $H$.

\subsection{The variational ansatz}

We choose the terms $a_k$ in Eq.~(\ref{VarEq}) in a repeating pattern and perform $S$ repetitions, which we call ``steps", 
In each step, we have three variational parameters, $\theta_h^b,\theta_v^b,\theta_U^b$, where $b=1,\ldots,S$. We set
\be
\label{Sansatz}
\Psi_T=\prod_{b=1}^S \Bigl( U_U(\frac{\theta_U^b}{2}) U_h({\theta_h^b})U_v(\theta_v^b) U_U(\frac{\theta_U^b}{2})\Bigr) \Psi_I,
\ee
where
$U_X(\theta)$ approximates $\exp(i \theta h_X)$ for $X\in \{U,h,v\}$ and the product is ordered by decreasing $b$; we say ``approximates" because
for $h_h$ is a sum of non-commuting terms and we therefore used a second-order Trotter-Suzuki method to implement $U_h$, applying the terms in sequence from left to right in each row and then reversing the order. This Trotterization helps to approximately preserve momentum, which is useful if $\Psi_I$ has the right momentum quantum numbers. 
For $U_U,U_v$, we implement the exponential exactly. 
Each step of the ansatz then is a second-order Trotter approximation to evolution under $H(b) \equiv \theta_U^b h_U+\theta_v^b h_v + \theta_h^b h_h$, (note that $[U_h,U_v]=0$; this property holds in general for free fermion hopping terms on any square lattice because $U_h$ and $U_v$ are diagonal operators in a basis for the Fock space obtained from a momentum basis for single-particle states).

The sequence length can be reduced by combining $U_U(\frac{\theta_U^{b+1}}{2}) U_U(\frac{\theta_U^b}{2}) = U_U(\frac{\theta_U^b+\theta_U^{b+1})}{2})$; other orderings, such as first applying odd terms and then applying even terms will slightly reduce the gate depth.

It is important to emphasize that we are {\it not} concerned with Trotter error other than for a desire to preserve quantum numbers such as momentum; while this choice of unitaries gives evolution similar to the evolution under $H(b)$, we are optimizing the parameters for evolution with these unitaries. If we instead did exact evolution under a time-varying Hamiltonians (which is possible for atomic quantum gases in optical lattices  \cite{toolbox}), we would instead optimize to a different optimum in the parameters. We expect that the difference in these evolutions can be largely absorbed into a small shift in the variational parameters.  Alternatively, one can think that instead of the terms $h_{a_i}$ in Eq.~(\ref{VarEq}) being chosen from the terms $h_U,h_v,h_h$ they correspond to the terms $h_U,h_v$, as well as all individual hopping terms that sum up to $h_h$, with all those hopping terms using the same parameter in a given step.

\subsection{Optimization with Exact Energies}
\label{sec:exact}

We first consider the case where we can exactly calculate energies on a classical computer and use a simple optimization procedure. Six points (a ``point" is a set of parameters) are chosen at random near the origin. For each point, we first follow a greedy noisy search, where we slightly perturb the values of the points, accepting whenever this reduces the energy, for a total of $150$ evaluations of the energy. We then use Powell's conjugate direction  \cite{powell} method until it converges.
After following this procedure for each of the six points, we keep the point whose energy is lowest at the end of the procedure, and we discard
the other five points. For the point we keep, we alternate greedy noisy search and Powell search until neither can find an improvement. Our greedy noisy search uses a simple algorithm to determine step size: every thirty trials we count the number of acceptances. If that number is large, the step size is increased; if the number is small, the step size is reduced.

We call this optimization algorithm the ``global variational" method as it involves optimizing all parameters simultaneously.
Using the global variational approach and an ansatz with $S=3$, we obtain energy errors of $2.0\times 10^{-8}$, $0.019$,$0.029$, $0.083$, and $0.59$ for $N=4$, 6, 8, 10, and 12. Increasing $S$ improves the error, but at the larger sizes convergence was slow and results varied greatly from run to run, suggesting that the minimization was getting stuck in local minima.
One clue to the origin of this difficulty is that in some cases the energy reduced but the overlap with the ground state also reduced. This is possible if it also reduces the amplitude of some highly excited state but increases the amplitude of a low excited state. 

To overcome this problem, we used an alternative procedure to find the optimum.  This procedure is inspired by annealing and we call it the ``annealed variational" method. Let $H_s=t h_H + t h_V + s U h_U$ interpolate from $H_0$ to $H_1$ by changing the coupling $U$.
We use the same ansatz (\ref{Sansatz}), but we use a different procedure to select a starting point for the search. We first do a single step ansatz using as $\Psi_I$ the ground state of $H_0$ and targetting $H_{1/S}$, using the same optimization as above, calling the resulting parameters $\theta_X^1$. We then use the $\Psi_T$ from that optimization as $\Psi_I$ for another single step targetting $H_{2/S}$, giving $\theta_X^2$. 
We continue, targetting $H_{3/S}, \ldots, $, using $\Psi_T$ from one step as $\Psi_I$ for the next. This then gives a trial state using $S$ steps for $H_1$ using parameters $\theta_X^b$ for $b=1,\ldots,S$. We call this state the result of sequential optimization.
We use those parameters as the starting point for a further search as before (i.e., a global variational search using the sequential optimization as the starting point), calling the resulting parameters the result of full optimization.  (An alternate method that we  tried that  in some cases worked better was to do a similar anneal, but to target $H_1$ on every step; thus, one would parameters for a single step ansatz targetting $H_1$; then use that to do another step again targetting $H_1$, and so on; this sometimes did better at a small number of steps and also did slightly better at the given number of steps on the $N=12$ site model.)


\begin{table}
\scalebox{0.9} {
\begin{tabular}{c | c | c | c | c || c | c | c|c|c}
$S$ & $\Delta E^s$ & $\Delta E^f$ &$P^s$ & $P^f$ & $\Delta E^s$ & $\Delta E^f$ &$P^s$ & $P^f$ \\ \hline
& 4 & & & & 6 & & & & \\
3 &  $0.24$ & $1.00 \times 10^{-8}$  & 0.7180  & 1.0000& $0.062$ & $0.033$ & 0.9821 &0.9903 \\
5 &  $0.20$ & $3.00 \times 10^{-8}$  & 0.76289 & 1.0000 & $0.034$ & $0.002$ & 0.9912 &0.9995 \\
7 &  $0.17$ & $2.00 \times 10^{-8}$  & 0.8021 & 1.0000 & $0.018$ & $0.00033$ & 0.9954 &0.9999  \\
9 &  $0.15$ & $7.00 \times 10^{-8}$  & 0.8275 & 1.0000 & $0.013$ & $0.00018$ & 0.9967 &1.0000  \\
11 & $0.13$ & $2.00 \times 10^{-8}$  & 0.8460 & 1.0000 & $0.012$ & $0.00011$ & 0.9970 &1.0000  \\
\hline

\hline
\hline
& 8 & & & & 10 & & & & \\
3 &   $0.1$ &   $0.033$  &  0.9790 & 0.9934   &$0.13  $&	$0.083 $ & $0.9217 $&	0.9374 \\
5 &   $0.042$ & $0.0046$  &  0.9906	& 0.9983&$0.087 $&	$0.041 $ & $0.9388 $&	0.9585 \\
7 &   $0.031$ &	$0.0030$  &  0.9930	& 0.9989&$0.066 $&	$0.022 $ & $0.9492 $&	0.9710 \\
9 &   $0.024$ &	$0.0013$  &  0.9947	& 0.9995&$0.053 $&	$0.014 $ & $0.9566 $&	0.9809 \\
11 &  $0.019$ &	$0.00089$  &  0.9960	& 0.9997&$0.042 $&	$0.012 $ & $0.9626 $&	0.9841 \\
13 &  $0.015$ &	$0.00038$  &  0.9968	& 0.9999&$0.036 $&	$0.0069 $ & $0.9660 $&	0.9929 \\
15 &  $0.013$ &	$0.00031$  &  0.9973	& 0.9999&$0.033 $&	$0.0052 $ & $0.9676 $&	0.9959 \\
17 &  $0.012$ &	$0.00022$  &  0.9976	& 0.9999&$0.032$&	$0.0032 $ & $0.9682 $&	0.9983 \\
19 &  $0.010$ &	$0.00027$  &  0.9978	& 0.9999&$0.032 $&	$0.0017 $ & $0.9688 $&	0.9993 \\

\hline
& $8^*$ & & & & 12 & & & & \\
3 & 0.66 &	$0.53$ & 0.3889	&0.5231& 0.69 &	$0.59$ & 0.3046	&0.4102 \\
5 & 0.56 &	$0.17$ & 0.4620	&0.8727& 0.59 &	$0.39$ & 0.3576	&0.6154 \\
7 & 0.49 &	$0.065$ & 0.5164	&0.9353& 0.54 &	$0.18$ & 0.3958	&0.8520 \\
9 & 0.44 &	$0.046$ & 0.5600	&0.9501& 0.51 &	$0.087$ & 0.4256	&0.9294 \\
11 &0.40 &	$0.032$ & 0.5969	&0.9609& 0.48 &	$0.054$ & 0.4483	&0.9541 \\
13 &0.36 &	$0.022$ & 0.6280	&0.9685& 0.46 &	$0.035$ & 0.4667	&0.9714 \\
15 &0.33 &	$0.017$ & 0.6551	&0.9829& 0.45 &	$0.025$ & 0.4836	&0.9790 \\
17 &0.31 &	$0.010$ & 0.6799	&0.9910& 0.43 &	$0.021$ & 0.4997	&0.9823 \\
19 &0.28 &	$0.0083$ & 0.7030	&0.9935& 0.42 &	$0.015$ & 0.5152	&0.9883 \\

\end{tabular}
}
\caption{Performance using the annealed variational method. Numbers $4,6,8,10,12$ indicate different values of $N$. $8^*$ is described in text.
The left-hand column is number of steps, $S$. Quantities $\Delta E^s,\Delta E^f$ indicate error in ground state energy after sequential and full optimization, respectively. Quantities $P^s,P^f$ indicate absolute squared ground state overlap in those two cases, respectively. }
\label{tableanneal}
\end{table}

The results of the annealed variational method are listed in Tab. \ref{tableanneal}. For $S=3, N>4$, this algorithm is seen to be only
marginally worse than the global variational method, but we found that it was faster at finding the optimum (the number of energy evaluations required depended on $N$, but the annealed variational method always required fewer than the original method, and in some cases required a factor of five times fewer evaluations).
The most important advantage of the annealed variational method is that it becomes significantly more accurate
than the global variational method for $S>3$, where now the error drops consistently with increasing $S$, without the convergence issues seen using the global variational method.
While generally larger $N$ is more difficult, $N=4$ shows worse performance than $N=6$ , 8, and 10 after sequential optimization and only shows better performance after full optimization, perhaps due to the degeneracy of the non-interacting Hamiltonian for $N=4$.

An additional model shown in the table is called ``$8^*$". This has eight sites with the same geometry as before, but we change the hopping strength in the horizontal direction to $1/\sqrt{2}$ and change the sign on the horizontal hoppings which are periodic, thus effectively inserting a $\pi$-flux for hopping around a loop.
The resulting single particle spectrum has $N/2-2$ states below the Fermi energy and four states at the Fermi energy, giving the same many-body ground state degeneracy as for $N=12$ sites.
The flux model and $N=12$ show very similar errors and are both distinctly more difficult than the others. Nonetheless, using only $33$ parameters we obtain over $95\%$ overlap and with $45$ parameters $97.9\%$ overlap for $N=12$, despite the Hilbert space having a much larger dimension: ${{12}\choose{6}}^2=853776$ at the given filling.
For $N=10$, a $93.7\%$ overlap is obtained with only $9$ parameters, in a $63504$-dimensional space (symmetries such as total spin slightly reduce this dimension).

\subsection{Inexact Optimization: Gate Count and Run Time}
\label{sec:inexact}

Next we consider the effects of sampling error assuming one measures individual terms in the Hamiltonian on a quantum computer, considering a tradeoff between runtime and accuracy.
Clearly, given a large enough number of samples, one can reduce the statistical error to the point that one can obtain the same accuracy as above.
However, for a smaller number of samples, it is useful to modify the optimization algorithm. The trouble is that a small change in a point leads to only a small change in energy, and it thus requires a large number of samples to determine whether or not the energy improves. 

We thus used a different algorithm and tested it for $N=8$. Starting with all parameters equal to zero we randomly order the parameters and try increasing or decreasing each parameter by a constant, looking for an improvement This is repeated (with parameters re-ordered again randomly) with slightly changed constants, until no further improvement is found. To determine if there is an improvement, we sample at the given points until the difference between the energies becomes twice the standard deviation (or until a maximum number of samples is done).
On an average over ten runs, we were able to obtain more than 98\% overlap with the ground state using $506$ different point evaluations and $8.5 \cdot 10^5$ average samples per point, for $4.3\times 10^7$ total samples.

Now we wish to estimate the number of gates required to obtain a single sample, including preparing the state $\Psi_I$, implementing the unitaries,
and finally measurement. $\Psi_I$ is a Slater determinant and can be prepared using Givens rotations  \cite{hubbard} (using fewer gates than the strategies in Ref.~\onlinecite{somma}). The unitaries $U_U,U_h,U_v$ can be implemented efficiently using Jordan-Wigner cancellation techniques  \cite{circuits}. Finally, for measurement, the terms in the Hamiltonian can be broken into at most four commuting sets. For $N=4n=8,12,\ldots$ these are the Hubbard terms, the vertical hopping terms, and two sets for the horizontal hopping terms. Slightly more complicated divisions of the hopping terms are needed for $N=4n+2=6,10,\ldots$, while a model with more horizontal rows will require five sets.

The cost of implementing the unitaries dominates the measurement cost.
We can measure $h_U$ with a cost (in gate count) that is almost identical to the cost to implement $U_U$; see, for example, the discussion
around Fig.~12 of Ref.~\onlinecite{hubbard} and references therein. Similarly, we can measure $h_v$ with cost almost identical to the cost to implement
$U_v$, while we can measure both sets of commuting terms in $h_h$ with cost almost identical to the cost to implement $U_h$.
Hence, since in a single run we only measure {\it one} of these four sets of terms (i.e., $h_U$ or $h_v$ or one of the two terms in $h_h$, assuming $N=4n$),
the measurement cost in a single run is roughly one-quarter the cost of a single step $ U_U(\frac{\theta_U^b}{2}) U_h({\theta_h^b})U_v(\theta_v^b) U_U(\frac{\theta_U^b}{2})$.
In general,
the cost of implementing $\prod_{b=1}^S \Bigl( U_U(\frac{\theta_U^b}{2}) U_h({\theta_h^b})U_v(\theta_v^b) U_U(\frac{\theta_U^b}{2})\Bigr)$
scales linearly with $S$.

The Slater determinant can be prepared using Givens rotations and the fast Fourier strategy  \cite{hubbard}, requiring on the order of $N \log_2(N)$ Givens rotations. These rotations will involve sites which are further separated than the nearest neighbor sites which appear in $U_h,U_v$, requiring longer Jordan-Wigner strings; ignoring the cost of the strings, the cost to execute these rotations will be
comparable to the cost to execute
$\log_2(N)$ rotations by $U_h$ or $U_v$. For fixed $S$, this would eventually dominate at large $N$, but we expect that $S$ will need to increase with $N$
too and that the dominant cost will remain the cost of implementing the steps. Consider the case of $N=8$, for example. The initial ground
state has two particles in the zero momentum state in the horizontal direction, one in the symmetric state in the vertical direction and one in the
anti-symmetric state. There are also two particles in momentum states $\pm \pi/2$ in the horizontal direction, in the symmetric state in the vertical direction.
We can prepare the ground state as follows. Label sites in the top row $1,2,3,4$ and label sites in the bottom row $5,6,7,8$ with $1$ directly above $5$, and so on. Initialize with
particles in sites $1,3,4,5$. Then, applying Givens rotations between pairs of sites $3,7$; $4,8$ to place particles initially in sites $3,4$ in the symmetric state in the vertical direction.
Next apply Givens rotations between pairs $1,2$; $5;6$ to bring the particle in site $1$ into a symmetric state between $1,2$ and the particle in site $5$ into
a symmetric state between $5,6$. Again apply Givens rotations between pairs $1,3$; $2,4$; $5,7$; $6,8$ so that now the particle initially in site $1$ is in a symmetric state between sites $1,2,3,4$ while the particle in site $5$ is in a symmetric state between sites $5,6,7,8$. Thus, we have successfully
occupied both states with zero momentum in the horizontal direction.
This same procedure in fact also produces the desired particles in momentum states $\pm \pi/2$ so it prepares the desired ground state.
This uses a total of $8$ Givens rotations. The gate count cost is comparable to that to implement $U_h$ and $U_v$.

For the case of quantum chemistry discussed below, the initial state is much simpler. It is a product state and so the cost to prepare is negligible compared to that to implement the unitaries.

A gate count
estimate shows that about $1000$ gates are required for $S=2$ at $N=8$ for a single run. This includes the cost to prepare $\Psi_I$, to implement the unitary rotations to create $\Psi_T$, and to perform measurement.
To measure all terms, this must be multiplied by $4$ as each run will give
a measurement of the terms in {\it one} of the four commuting sets of terms.
To give a rough estimate, if we ignore the possibility of parallelizing the circuits and assume a gate time of $1\mu s$ the total time would be $4.3\times10^7\times4\times1000\times10^{-6}$ seconds, or roughly $47$ hours. This could be further improved using several devices in parallel to sample.
Note that any given run requires only about $1000$ gates, and thus poses only moderate demands on gate fidelities.

\section{ Quantum Chemistry}
\subsection{The electronic structure Hamiltonian}
We finally apply the method to three problems in quantum chemistry, where there are more terms than in the Hubbard model. Using the Born-Oppenheimer approximation by assuming that the nuclei behave classically and are localized, the Hamiltonian for the electronic degrees in second quantized form
\begin{equation}
H=\sum_{pq} h_{pq} c^\dagger_p c_q+\sum_{pqrs} h_{pqrs} c^\dagger_p c^\dagger_q c_r c_s
\label{eq:qc}
\end{equation}
has the most general form for two-body interactions, due to the long range nature of the Coulomb interaction. The indices $p$, $q$, $r$ and $s$ refer to spin-orbitals, combining the orbital and spin index. 

As most basis sets used in quantum chemistry are non-orthogonal we follow the standard procedure  \cite{whitfield} of first performing a (classical) Hartree-Fock calculation and then use the Hartree-Fock orbitals as an orthogonal basis set. On classical computers we can only simulate quantum algorithms for very small basis molecules. 

Specifically we consider HeH$^+$ which has $two$ electrons, using a P321 basis with $N_{\rm SO}=8$ spin orbitals, H$_2$O which has $10$ electrons, in an STO-6G basis with $N_{\rm SO}=14$, and BeH$_2$ which has $6$ electrons in a basis with $N_{\rm SO}=14$. We use the Psi4  \cite{Psi4} quantum chemistry package to perform the Hartree Fock calculation and compute the matrix elements of the Hamiltonian (\ref{eq:qc}). We also consider artificial hydrogen chains H$_N$, where we space hydrgen atoms along a line with distances of either 0.4614{\AA} or 2\AA.  We used the global variational approach to optimize parameters for HeH$^+$, H$_2$O, and BeH$_2$ and we used the annealed variational approach to optimize H$_N$.   All numbers reported for all methods are the best result obtained from three runs; due to randomness in the search the results differ slightly between runs.

\subsection{Variational ansatz for quantum chemistry}

We considered two differents ansatzes for quantum chemistry.  In the first ansatz, used for HeH$^+$, H$_2$O, BeH$_2$,
instead of using $\mathcal(O)(N_{\rm SO}^4)$ variational parameters in our ansatz, we group the matrix elements into three terms
$H=H_{\rm diag}+H_{\rm hop}+H_{\rm ex}$, where
\begin{equation}
H_{\rm diag}=\sum_{p} \epsilon_p c^\dagger_p c_p + \sum_{p,q} h_{pqqp} c^\dagger_p c_p c^\dagger_q c_q
\end{equation}
is a sum of diagonal terms,
\begin{equation}
H_{\rm hop}=\sum_{pq} h_{pq} c^\dagger_p c_q + \sum_{prq} h_{prrq} c^\dagger_p c_q c^\dagger_r c_r
\end{equation}
contains normal and correlated hopping terms and
\begin{equation}
H_{\rm ex}=\sum_{pqrs} h_{pqrs} c^\dagger_p c^\dagger_q c_r c_s
\end{equation}
 for $p,q,r,s$ all distinct contains all other exchange terms.

\begin{table}
\begin{tabular}{c|c|c||c|c||c|c}
$S$ & $\Delta E$ [mHa] & $P$ & $\Delta E$ [mHa] & $P$ &$\Delta E$ [mHa] & $P$\\
\hline
& HeH$^+$ && H$_2$O&& BeH$_2$ & \\
1 & $7.8$ & 0.9970 & $23$ & 0.9886 & $22 $ & 0.9825 \\
2 & $1.7$ & 0.9992 & $9.5$ & 0.9950 & $6.6$ & 0.9935\\
3 & $0.59$ & 0.9998 & $7.6$ & 0.9955 & $6.3$ & 0.9937\\
4 & $0.26$ & 0.9999 & $6.8$ & 0.9959 & $5.8$ & 0.9939\\
5 & $0.088$ & 0.9999 & $3.2$ & 0.9980 & $4.23$ & 0.9954\\
6 & $0.14$ & 0.9999 & $3.1$ & 0.9982 & $1.85$ & 0.9977\\
\end{tabular}
\caption{Variational Method applied to HeH$^+$ and H$_2$O. $\Delta E$ is the error in the ground state energy, $P$ is absolute squared overlap with ground state. Note that the energy is slightly worse for HeH$^+$ at $S=6$ than at $S=5$; this indicates that the optimization algorithm did not find the true minimum in this case.}
\label{tableqc}
\end{table}

As our final ansatz we use
\begin{eqnarray}
\label{chemansatz}
\Psi_T 
&=&\prod_{b=1}^S  \Bigl( U_{\rm ex}(\frac{\theta^b_{\rm ex}}{2}) U_{\rm hop}(\frac{\theta_{\rm hop}^b}{2}) \\ \nonumber && \times U_{\rm diag}(\theta_{\rm diag}^b) U_{\rm hop}(\frac{\theta_{\rm hop}^b}{2}) U_{\rm ex}(\frac{\theta^b_{\rm ex}}{2}) \Bigr) \Psi_I,
 \end{eqnarray}
where $U_X(\theta)$ approximates $\exp(i \theta H_X)$ for $X \in \{{\rm ex},{\rm hop}, {\rm diag}\}$, up to Trotterization error.
The initial state $\Psi_I$ is chosen to be the ground state of $H_{\rm diag}$ and the basis is a Hartree-Fock basis so that
$H_{\rm hop} \Psi_I =0$.  The unitary $U_{\rm diag}(\theta)=\exp(i \theta H_{\rm diag})$ can be implemented
exactly since all terms commute. We implemented the unitary $U_{\rm hop}(\theta)$ using an ``interleaved" term ordering  \cite{circuits}, approximating it as a product 
$U_{\rm hop}(\theta)=\prod_{p<q}U_{pq}(\theta)$ in arbitrary order, with 
\begin{equation}
U_{pq}(\theta)=\exp\left[i \theta\left (h_{pq} c^\dagger_p c_q +\sum_r h_{prrq} c^\dagger_p c^\dagger_r c_r c_q +h.c.\right)\right],
\end{equation} 
where all terms in the exponential commute with each other.
The term $U_{\rm hop}$ appears twice in Eq.~(\ref{chemansatz}) for each step; in a slight abuse of notation, the order of terms was reversed in the two different applications. 
The unitary $U_{\rm ex}(\theta)$ was approximated as a product over factors
\begin{equation}
U_{pqrs}(\theta)=\exp\left[i \theta\left (h_{pqrs} c^\dagger_p c^\dagger_r qc_r c_s +h.c.\right)\right],
\end{equation} 
in an arbitrary order. 
The term $U_{\rm ex}$ also appears twice in Eq.~(\ref{chemansatz}) for each step; in another slight abuse of notation, the order of terms was reversed in the two different applications.
Thus, in each step the unitary applied is a second-order Trotter approximation to evolution under $H(b)\equiv \theta^b_{\rm ex} H_{\rm ex}) + \theta^b_{\rm hop} H_{\rm hop} + \theta^b_{\rm diag} H_{\rm diag}$.

The second ansatz was used for the H$_N$ chains.  In this case we used four parameters.  We did this by further dividing $H_{ex}$ into a sum of two
terms
$H_{\rm ex}=H_{\rm o} + H_{\rm rest}$, where $H_{\rm o}$ includes the terms in $H_{\rm ex}$ which do not annihilate the 
Hartree-Fock state (the ``o" indicates that this is based on occupancy of the orbital in the Hartree-Fock state) and $H_{\rm rest}$ includes the remaining terms.
We then used separate parameters to control $H_{\rm o},H_{\rm rest}$.  Thus, the final ansatz was
\begin{eqnarray}
\label{chemansatz4}
\Psi_T 
&=&\prod_{b=1}^S  \Bigl( U_{\rm o}(\frac{\theta^b_{\rm o}}{2})  U_{\rm rest}(\frac{\theta^b_{\rm rest}}{2}) U_{\rm hop}(\frac{\theta_{\rm hop}^b}{2}) \\ \nonumber && \times U_{\rm diag}(\theta_{\rm diag}^b) U_{\rm hop}(\frac{\theta_{\rm hop}^b}{2}) U_{\rm rest}(\frac{\theta^b_{\rm rest}}{2}) U_{\rm o}(\frac{\theta^b_{\rm o}}{2}) \Bigr) \Psi_I,
 \end{eqnarray}

Our results for the three benchmark molecules using various numbers of steps $S$ are shown in Table \ref{tableqc}.

\subsection{Unitary Coupled Cluster Ansatz}
We have also performed simulations using UCC  \cite{ucc} for comparison as well as variants. UCC
method fixes $\Psi_T=\exp(T) \Psi_I$, for $T$ an anti-Hermitian operator. The variational parameters are contained in the choice of $T$.
Ref.~\onlinecite{aag} proposes using a Trotter-Suzuki method to implement the transformation $\exp(T)$. This increases the depth of the circuit, which we wish to avoid. We have found in our simulations that more accurate results are in fact obtained using large Trotter steps, with the most accurate results obtained using between two and four second-order Trotter-Suzuki steps; all results reported below are for two such steps. The method thus differs from the usual UCC, although we continue to refer to it as such. We used similar optimization procedures as before, exactly evaluating (to numerical error) the energy as a function of parameters.

We set
\begin{equation}
\label{Tdef}
T=\sum_{p<q} (T_{pq} c^\dagger_p c_q - h.c.) + \sum_{p<q,r<s} (T_{pqrs} c^\dagger_p c^\dagger_q c_r c_s - h.c.),
\end{equation}
 using all
possible quadratic and quartic terms which are compatbile with the symmetries of the system. 
We considered two cases, taking the parameters $T_{pq}$, $T_{pqrs}$ either all real or all imaginary. More accurate results were found using real choices; this is likely due to the fact that the wavefunction is real. More flexibility could be obtained by using general complex values, but at the cost of a further increase in parameters. In the real case, we can drop diagonal terms such as $c^\dagger_p c_p$.   Thus, all runs reported are for the choice of real parameters.

Further, in the UCC method, one considers only terms in which all creation operators act on orbitals which are unoccupied in the Hartree-Fock state and all annihilation operators act on
orbitals which are occupied, or vice-versa as required by Hermiticity; equivalently, these are the terms which do not annihilate the Hartree-Fock state.  In our numerics, we considered three alternative possibilities for a total of four possible choices of terms
to keep.  One choice we call RAA, in which {\it all} terms are kept (the ``A"s refer to keeping all quadratic and all quartic terms, while the ``R" refers to $T$ being real).
Another choice we call ROO, in which now we keep only quadratic and quartic terms which do not annihilate the Hartree-Fock state (the ``O" refers to ``occupation", as whether
or not the term is kept depends upon the occupation number).  The choice ROO is precisely the same as UCC.  The next choice we term RNO, where {\it no} quadratic terms are kept and quartic terms are kept only if they do not annihilate the Hartree-Fock state (the ``N" refers to ``none"); this choice has the fewest terms.  The last choice is RAO, where
all quadratic terms are kept and only quartic terms are kept if they do not annihilate the Hartree-Fock state.  Thus, in decreasing order of number of terms kept, they are RAA, RAO, ROO, RNO.  The variants can be collectively termed Rxx.

For molecules, to find terms compatible with symmetries, we included all terms which had nonzero coefficients in the Hamiltonian (this should work for small
molecules, while for large molecules some terms compatible with symmetry will have
a small enough coefficient that they get truncated).
The results are shown in Table \ref{tableRxx}.
Note that for HeH$^+$, H$_2$O, all methods are able to find the ground state to very high accuracy.  This may be due to the large number of parameters compared to Hilbert space dimension.  For RAA, for HeH$^+$, there are $192$ parameters in the imaginary case (almost as many in the real case) compared to a Hilbert space dimension of $64$ with the given number of up and down electrons (possibly further reduced by symmetry) while for H$_2$0 there are $595$ parameters compared to a Hilbert space dimension of $441$.
This is an obstacle that we encounter when trying to simulate small molecules on a classical computer. The number of variational parameters in this method grows as $N_{\rm SO}^4$ while the Hilbert space dimension grows exponentially with $N_{\rm SO}$ at fixed filling fraction; eventually the exponential growth beats the polynomial (indeed, this is the whole reason for interest in a quantum computer), but since the number of parameters grows at such a high power of $N_{\rm SO}$,
one needs fairly large systems to see this.  The other methods have smaller numbers of parameters; some slight accuracy loss can be seen in RNO.

The reduction in the number of terms when going from RAA to RNO or ROO depends upon the specific molecule.  At half-filling, for molecules with $N_{SO}$ spin-orbitals with $N_{SO}>>1$, the number of quartic in RAA is proportional to $N_{SO}^4$ and will be much larger than the number of quadratic terms.
Roughly $1/8$-th of these terms will be retained in RAO, RNO, or ROO; to see this, note that a term is retained if $p,q$ both correspond to unoccupied orbitals
and $r,s$ both to occupied (which occurs with probability $1/2^4=1/16$ if one selects $p,q,r,s$ randomly at half-filling) or $p,q$ are both occupied and
$r,s$ are both unoccupied (which also occurs with probability $1/16$).
This leads thus to a constant factor gain; away from half-filling the constant factor improvement becomes larger.

For BeH$_2$, the Hilbert space dimension is $1225$, compared to $595$ terms for RAA and so the Hilbert state dimension is smaller than the number of terms, but still comparable. 
For BeH$_2$, 
RAA is able to achieve an energy error of $0.157$ mHa, at the cost of over $2 \times 10^5$ energy evaluations.  A smaller number of evaluations leads to reduced accuracy; the accuracy improves most rapidly up to roughly $2 \times 10^4$ evaluations with an error of roughly $0.5$ mHa at that point.
 In contrast, the Hamiltonian variational method used between $5000-10000$ evaluations for the data above.
For a given number of evaluations in the optimization procedure, in general RAA had slightly improved energy than the methods with fewer terms.

\begin{table}
\begin{tabular}{c|c|c|c}
& HeH$^+$ &H$_2$O& BeH$_2$  \\
\hline
RAA & $3.2 \times 10{-4}$ & $3.3 \times 10^{-2}$  & $0.16 $  \\
RAO & $5.3 \times 10^{-3}$  & $4.4 \times 10^{-2}$  & $0.41 $\\
ROO & $1.3 \times 10{-2}$  & $5.9 \times 10^{-2}$ & $0.42 $ \\
RNO & $0.42$& $0.36$ & $0.66$ \\
\end{tabular}
\caption{Rxx methods applied to various small molecules.  Table shows energy error in mHa.  Note that ROO is the same as UCC.}
\label{tableqcRxx}
\end{table}

\subsection{Hydrogen chains}

Real applications will be concerned with molecules with larger $N_{\rm SO}$. In order to access this regime we turn to a different system, H$_N$ chains, varying the number of atoms from $N=2$ to $N=10$;  this allows us to test scaling by considering a sequence of different system sizes, all at fixed filling fraction and with
similar couplings.
For $N=2$, 4 and 6 the number of terms in RAA exceeds the Hilbert space dimension, while for $N=8$ there are $2964$ terms with a Hilbert space dimension of $4900$ and for $N=10$ there are $7230$ terms with a Hilbert space dimension of $N=63504$.
We performed simulations at separations of either 0.4614{\AA} or 2{\AA} between the atoms. For the smaller spacings, there is a much higher overlap between the true ground state and the Hartree-Fock ground state, while for the larger spacing, the overlap becomes much smaller; this is similar to the Hubbard chain at small $U/t$ compared to large $U/t$.

 The results are shown in Table \ref{tableHn}.   In general, for larger $N$, the Hamiltonian variational method is able to obtain significantly higher overlap (the use of four parameters improves this; with only three parameters, a larger $S$ is required to attain the same accuracy).

\begin{table}
\begin{tabular}{c|c|c||c|c||c|c}
 & $\Delta E$ [mHa] & $P$ & $\Delta E$ [mHa] & $P$ &$\Delta E$ [mHa] & $P$\\
\hline
& H$_6$ && H$_8$&& H$_{10}$ & \\
RAA & $0.4$ & 0.995 & $51$ & 0.786 & $152 $ & 0.578 \\
RAO & $8.03$ & 0.976 & $54$ & 0.797 & $89$ & 0.674\\
ROO & $10$ & 0.979 & $54$ & 0.787 & $80$ & 0.697\\
RNO & $13$ & 0.974 & $50$ & 0.817 & $69$ & 0.746\\
H Var $3$ & $21$ & 0.947 & $35$ & 0.904 & $51$ & 0.892\\
H Var $6$ & $3.0$ & 0.982 & $7.3$ & 0.959 & $11$ & 0.960\\
H Var $9$ & $0.97$ & 0.984 & $2.6$ & 0.965 & $3.6$ & 0.967\\
\end{tabular}
\caption{Variational methods applied to hydrogen chains.  Left-hand column gives method.  Rxx refers to Rxx methods with ROO=UCC while H Var $3,6,9$ refer to Hamiltonian variational with $S=3,6,9$.}
\label{tableHn}
\end{table}

 For H$_6$, increasing the number of terms in Rxx leads to improved performance: RAA outperforms RAO which outperforms ROO which outperforms RNO on measures of energy.  However, for H$_8$, H$_{10}$, the reverse is true, with
RNO having the highest overlap of any Rxx method.
When we consider the energy accuracy as a function of number of evaluations, we find that at a given number of evaluations RAA generally performs better (often by only a small amount).  However, the methods with fewer parameters are often able to continue the optimization out for more steps before getting stuck.  We do not fully understand the mechanism for this, but one possibility is that once the optimization routine gets very close to a local minimum,
it is likely that motion in a random direction will hurt the energy.  Methods with more parameters are more likely to move in an incorrect direction simply due to the increased dimensionality of the search space and
hence will be more likely to fail to find an improvement once very close to the minimum.
One possible improvement then would be to run RNO until it found its optimum; then, take that endpoint as a starting point for an ROO or RAA search; this requires investigation.

An interesting further feature of the convergence is that the improvement in energy is rather rapid initially and then transitions to a routine with much
slower improvement.  For example, for H$_{10}$, the energy for the various Rxx versions has an error of $\sim 0.2$ after roughly $5\times 10^4$ evaluations, followed by a very gradual improvement, with the RNO numbers above occuring after over $2.5 \times 10^5$ evaluations.

As $N$ increases, we find that the annealed variational
approach gets increasingly better than the UCC; the UCC shows good performance at small sizes, perhaps due to the large number of parameters compared to the Hilbert space dimension, but gets increasingly less accurate at large sizes and is more difficult to optimize.
For the smaller spacing, Rxx performed relatively better but a similar trend was found.

A more flexible method for quantum chemistry is to use multiple steps (as in the Hamiltonian variational approach) but allow
every term to have a separate parameters (as in UCC);  we call such method TRxx for ``time-dependent Rxx", where the ``time" refers to different steps. Optimizing all these parameters may become difficult.
Note that for Hubbard, the unitary coupled cluster method requires a large increase in circuit depth compared to the Hamiltonian variational approach due to all the additional terms in $T$ and hence it is likely to not be suitable for that context unless particularly useful choices of a small number of terms in $T$ can be found. The Hamiltonian variational approach can be regarded as a useful choice of which terms to take and how to vary them.

\section{Resource estimates for practical applications}

We want to end with a discussion about resource requirements for practical applications that might go beyond what can currently be done classically.

\subsection{Hubbard model}
For the Hubbard model interesting results can be obtained for lattice sizes beyond $10\times10$, and thus with $N\ge100$ sites, which requires about 200 qubits (or slightly more for ancillas if we want to parallelize the circuits). As the circuits to implement the various terms in the Hamiltonian can be efficiently parallelized \cite{hubbard}, the parallel circuit depth will not substantially increase except for the potential need to go to a larger number of steps $S$. In order to be competitive to the best classical variational wave functions and distinguish competing ground states (such as stripe oder versus uniform superconducting ground states) we need to aim for an error of $\epsilon\approx10^{-3}t$ in the energy {\em per site} \cite{corboz}.

Assuming a variance of order $1$ for the measurement of the hopping terms $c^\dagger_{i,\sigma} c_{j,\sigma}+c^\dagger_{j,\sigma} c_{i,\sigma}$ and a reduced variance of order $1/U$ for the double occupancy term $c^\dagger_{i,\uparrow} c_{i,\uparrow} c^\dagger_{i,\downarrow} c_{i,\downarrow}$ (due to a suppression of double occupancy to about $t/U$ for large $U$) we get an error estimate of
\begin{equation}
\epsilon^2 \approx U^2 \frac{t/U}{M N } + 4 t^2 \frac{1}{M N},
\end{equation}
assuming $M$ measurements for each of five terms (double occupancy, horizontal and vertical hopping for each of the spins) and using that we can perform $N$ measurements in parallel in each of the runs. This is consistent with the number of samples that was found to be necessary comparing this to the values measures for small systems in Sec. \ref{sec:inexact}. Using relevant values of $t=1$, $U=8$, $N=100$, we obtain $M \approx 12/{\epsilon^2 N}\approx 120,000$ samples for each of the five terms, or about a total of 600,000 samples per energy evaluation. Again assuming gate times of order $1\mu s$, as in  Sec. \ref{sec:inexact}, the total estimated times needed to achieve convergence remains of the order of days when parallelizing the circuits, even considering that larger $S$ and more energy evaluations might be required. While not trivial on a small quantum computer, this is potentially practical in the not too distant future. 

\subsection{Quantum chemistry}

For quantum chemistry applications the required resources will be more demanding. Writing the Hamiltonian as
$
H=\sum_{i=1}^{N_{\rm terms}} h_i {\cal O}_i,
$
where $N_{\rm terms}$ is the number of terms, and the numbers $h_i$  the coefficients of the terms ${\cal O}_i$ we obtain for the error
\begin{equation}
\epsilon^2 = \sum_i \frac{|h_i|^2 {\rm Var}({\cal O}_i)}{M_i},
\end{equation}
assuming $M_i$ measurements of each of the operators ${\cal O}_i$, where ${\rm Var}(\ldots)$ denotes the variance in the measurement of the operator for the given trial state. Minimizing the error for a total number of measurements $M$, we choose $M_i\propto |h_i|$, and bounding the variances by  ${\rm Var}({\cal O}_i)\le 1$ we get
\begin{equation}
M \approx  \frac {\left(\sum_i |h_i|\right)^2}{\epsilon^2}
\end{equation}
As the variances of the (large) diagonal terms $\epsilon_p$ and $h_{pqqp}$ will be small for orbitals where the occupation number is close to $1$ or to $0$ (this holds for all orbitals in the molecules studied above), we will drop these terms from the error estimates and consider just the off-diagonal terms in the following order of magnitude estimate.
Some of the off-diagonal terms have small variance.  For example, any term $h_{pq}$ in which $p,q$ both have occupation number
close to $1$ or close to $0$ will have small variance.  However many off-diagonal terms have large variance; for example, a term $h_{pq}$ where $p$ has occupation number close to $1$ and $q$ has occupation number close to $0$ has large variance.
Hence, as a rough estimate, we treat the variance of the off-diagonal as a constant of order unity.

We find for the sums of matrix elements $\sum_i |h_i|$ the values 11.3Ha for HeH$^+$, $12.3$Ha for BeH$_2$ and 36Ha for H$_2$O, which results in $10^8$ to $10^9$ required samples for each energy evaluation to achieve an error of 1mHa. This is about a factor 1000 larger than for the Hubbard model, which already needed on the order of a few days to be optimized at an assumed $1\mu s$ gate time and measurement time. Considering additionally that the circuits are more complex, this will require a massively parallel cluster of quantum computers to perform the simulations for small molecules in a reasonable time.

Moving from these toy problems that can easily be simulated classically to a more challenging problem, such as Fe$_2$S$_2$ in 
a small single particle basis (STO-3G with $N=112$ spin-orbitals) we have $\sum_i |h_i|\approx 3$kHa, thus requiring about $10^{13}$ samples per energy evaluation. Even assuming that still only about $10^6$ energy evaluations could be sufficient to optimize the ansatz, the required number of $10^{19}$ total samples is too large for the variational algorithm to be practical in this form. For this model, a single instance of the Hamiltonian circuit is on the order of $2 \times 10^{8}$ gate executions, which is an estimate of the cost to prepare a sample using either a UCC method truncated at all four fermi terms or the Hamiltonian variational method.
This amounts to a total of $10^{26}$ gate operations assuming $20-30$ commuting terms are measured on each sample.  This estimate assumed that all of the energy evaluations in the optimization were conducted at the same accuracy as the final estimate; potentially some of the earlier energy evaluations could be conducted at lower energy accuracy, but following a noisy search to the minimum may require that many steps of the evaluation be conducted at {\it higher} accuracy than the final evaluation, since we have found
that often only a very small improvement in energy is obtained on each step of the search.

\section{Discussion}
We have presented a method of constructing variational states for arbitrary Hamiltonians, and tested it numerically. Preparation of quantum states using short quantum circuits is likely the lowest hanging fruit for small quantum computers to achieve quantum supremacy, and outperforming classical computers. No expensive quantum error correction may be needed if the gate fidelity is high enough to allow a short quantum circuit to succeed in preparing a state.

We find that a modest number of parameters enables us to obtain a very large overlap with the ground state, helping reduce the number of evaluations required to obtain good values of the parameters. Of course, the minimum possible circuit depth is attained simply by declaring the class of variational states to be ``all states that can be constructed with unitary quantum circuits of given depth". However, this would require an impractical optimization. The method here allows
us a flexible class of states with a small number of parameters, with the circuits chosen from the terms in the Hamiltonian.  We find that for larger systems with stronger interactions, this approach is able to obtain significantly higher overlap than ansatz wave functions based on the unitary coupled cluster formalism or its Rxx variants. 

Although the state preparation is fast, the number of measurements required to estimate the energy with sufficient accuracy is a huge challenge. While the estimates for interesting quantum chemistry applications are astronomical and the variational algorithm for these problems thus impractical in its current form, the demands for the Hubbard model are less challenging. There, the smaller number of terms,  limited energy range and translation invariance help reduce the required number of measurements to make a quantum variational approach potentially practical, although still demanding.

 We have also given a detailed numerical simulation of UCC and Rxx for a variety of systems, obtaining better understanding of its convergence properties.  Further, we have shown that small Trotter numbers suffice, reducing circuit depth for this algorithm.

A less demanding application may be to use these ansatz wave functions for state preparation, which uses shorter circuits than would be needed if one instead prepared the state by adiabatic preparation. We have attempted to compare this by optimizing
the parameters in an anneal for the Hubbard model: we consider an anneal using a linear path from the model at $U=0$ to the final model at $U=2$. This linear path was implemented by discrete time steps ${\rm d}t$ on the quantum computer, evolving using a second order Trotter-Suzuki for each time step (hence, the individual rotations in this Trotter-Suzuki evolve by time step ${\rm d}t/2$. Thus,
there are two parameters that quantify the anneal: the total annealing time $T$, and the time step ${\rm d}t$. The ratio $T/{\rm d}t$ is the number of 
second order Trotter-Suzuki steps that must be implemented. We optimized these two parameters separately to obtain the minimum ratio that 
achieved $0.99$ or higher overlap with the ground state. For the $N=8$ model, this minimum was $8$, to be compared to $S=3$ to attain this accuracy using the variational method. For $N=12$, the minimum was $640$, to be compared to $S=19$ using the variational method (the table shows only $0.9883$ overlap for $S=19$, but slightly higher overlap was obtained by targetting $H_1$ on every step in this case). We expect that larger system sizes may lead to even more significant gains in depth.

This state preparation may be useful for measuring properties of the states, in particular when combined with the quadratic speedup of Ref.~\onlinecite{hubbard}.  Additionally, this state preparation may be useful for annealing to even larger states.  For example, if we construct a short circuit to create a Hubbard model ground state on $12$ sites (and indeed, one outcome of this work is that we have been able to find such a circuit using a classical computer), then one could create two or more copies of this ground state and then anneal from the Hubbard Hamiltonian describing $24$ sites partitioned into $2$ decoupled sets of $12$ sites to the Hubbard Hamiltonian for $24$ coupled sites.  This is an example of the ``joining" procedure of Ref.~\onlinecite{hubbard}, using the variational method to simplify the creation of the decoupled systems.

We optimized the parameters by a search procedure which treated the energy as a function of parameters of a black-box function to be optimized.
It is in fact possible to also determine the derivative of this function using a quantum circuit, but this requires roughly doubling the depth of the circuit. It is not clear whether or not access to the derivative would improve the optimization.

One might further speculate whether optimizing over parameters by searching would be useful for the optimization problems of Refs.~\onlinecite{farhi1,farhi2}, improving over the performance of an adiabatic algorithm. Further numerical work may help answer this.

\acknowledgements
 We thank P. Love, J. McClean, and A. Aspuru-Guzik for discussion and useful explanations of UCC as in Refs.~\onlinecite{aag,ucc}.
This work was supported by Microsoft Research. MH and MT acknowledge hospitality of the Aspen Center for Physics, supported by NSF grant PHY-1066293.


\begin{thebibliography}{99}
\bibitem{mps} U. Schollwoeck, Ann. Phys. {\bf 326}, 96 (2011).

\bibitem{mera} G. Vidal,, Phys. Rev. Lett. {\bf 101}, 110501 (2008).

\bibitem{peps} F. Verstraete and J. I. Criac, cond-mat/0407066 (2004).

\bibitem{pepsprep1} M. Schwarz, K. Temme, and F. Verstraete, Phys. Rev. Lett. {\bf 108}, 110502 (2012).

\bibitem{pepsprep2} M. Schwarz, T. S. Cubitt, K. Temme, F. Verstraete, and D. Perez-Garcia, Physical Review A {\bf 88}, 032321 (2013).


\bibitem{schuch} N. Schuch, M. M. Wolf, F. Verstraete, and J. I. Cirac, Phys. Rev. Lett. 98, 140506 (2007).

\bibitem{univary} C. M. Dawson, J. Eisert, and T. J. Osborne,
Phys. Rev. Lett. {\bf 100}, 130501 (2008).

\bibitem{aag} A. Peruzzo et. al., Nature Comm. {\bf 5}, 4213 (2014).

\bibitem{ucc} A. Taube and R. J. Bartlett, Int. J. Quant. Chem. {\bf 106}, 3393–3401 (2006).

\bibitem{farhi1} E. Farhi, J. Goldstone, and S. Gutmann, arXiv:1411.4028.

\bibitem{farhi2} E. Farhi, J. Goldstone, and S. Gutmann, arXiv:1412.6062.

\bibitem{hubbard} D. Wecker, M. B. Hastings, N. Wiebe, B. K. Clark, C. Nayak, and M. Troyer, arXiv:1506.05135.

\bibitem{toolbox} D. Jaksch and P. Zoller, Annals of Physics, {\bf 315} 52 (2005).

\bibitem{powell} M. J. D. Powell, Computer Journal {\bf 7}, 155 (1964).

\bibitem{somma} R. Somma, G. Ortiz, J. E. Gubernatis, E. Knill, and R. Laflamme, Phys. Rev. A {\bf 65}, 042323 (2002).

\bibitem{circuits} M. B. Hastings, D. Wecker, B. Bauer, and M. Troyer, QIC {\bf 15}, 1 (2015).

\bibitem{whitfield} J. D. Whitfield, J. Biamonte, and A. Aspuru-Guzik, Molecular Physics {\bf 109}, 735 (2011).

\bibitem{Psi4} ``Psi4: An open-source ab initio electronic structure program," J. M. Turney, A. C. Simmonett, R. M. Parrish, E. G. Hohenstein, F. Evangelista, J. T. Fermann, B. J. Mintz, L. A. Burns, J. J. Wilke, M. L. Abrams, N. J. Russ, M. L. Leininger, C. L. Janssen, E. T. Seidl, W. D. Allen, H. F. Schaefer, R. A. King, E. F. Valeev, C. D. Sherrill, and T. D. Crawford, WIREs Comput. Mol. Sci. {\bf 2}, 556 (2012). (doi: 10.1002/wcms.93).

\bibitem{corboz} P. Corboz, T. M. Rice, and M. Troyer, Phys. Rev. Lett. {\bf 113}, 046402 (2014).

\end{thebibliography}
\end{document}